\documentclass[aps,prl,twocolumn,groupedaddress,floats,showpacs]{revtex4}
\usepackage{latexsym}
\usepackage{dcolumn}
\usepackage[dvips]{graphicx}
\usepackage{amssymb}
\usepackage{graphics}
\usepackage{amsmath}
\usepackage{epsf}

\newcommand{\vq}{\vec q}
\renewcommand{\vr}{\vec r}

\renewcommand{\vec}[1]{\mathbf{#1}}

\begin{document}

\author{Shaffique Adam$^{1}$, Piet W. Brouwer$^{2}$, and S. Das Sarma$^{1}$}
\affiliation{$^1$Condensed Matter Theory Center, Department of Physics,
University of Maryland, College Park, MD 20742-4111}
\affiliation{$^2$Laboratory of Atomic and Solid State Physics,
Cornell University, Ithaca, NY 14853-2501}

\title{Crossover from quantum to Boltzmann transport in graphene}

\date{\today}
\begin{abstract}
We compare a fully quantum mechanical numerical calculation of the
conductivity of graphene to the semiclassical Boltzmann theory.
Considering a disorder potential that is smooth on the scale of the
lattice spacing, we find quantitative agreement between the two
approaches away from the Dirac point. At the Dirac point the two
theories are incompatible at weak disorder, although they may be
compatible for strong disorder.  Our numerical calculations provide a
quantitative description of the full crossover between the quantum and
semiclassical graphene transport regimes.  
\end{abstract}
\pacs{73.23.-b,81.05.Uw,72.10.-d,73.40.-c}
\maketitle
 
Arguably, one of the most intriguing properties of graphene transport
is the non-vanishing ``minimum conductivity'' at the Dirac point. The
carrier density $n$ in these single monatomic sheets of carbon can be
continuously tuned from electron-like carriers for large positive gate
bias to hole-like carriers for negative
bias~\cite{kn:novoselov2005}. 
The physics close to zero
carrier density (also called the intrinsic or Dirac region), is now
understood to be dominated by the inhomogeneous situation where the
local potential fluctuates around zero, breaking the landscape into
puddles of electrons 
and holes~\cite{kn:katsnelson2006b}.
In the literature, two separate pictures have emerged to understand
 the value of this minimum conductivity.  The first picture expands
 around the universal value for the minimum conductivity $\sigma_{\rm
 min} = 4e^2/(\pi h)$ for clean
 graphene~\cite{kn:fradkin1986}
and argues that the presence of
 potential fluctuations smooth on the scale of the graphene lattice
 spacing {\em increases} the conductivity through quantum interference
 effects~\cite{kn:suzuura2002,kn:bardarson2007,kn:schuessler2008,kn:tworzydlo2008}.
 The second picture extrapolates to the Dirac region from the
 high-density limit, where the conductivity for charged impurity
 scattering can be calculated using the semiclassical Boltzmann
 theory. This approach has been further refined near the Dirac point
 by positing that the system acquires an effective carrier density
 $n^*$ calculated from the rms density fluctuations (associated with
 the electron-hole puddles) about the Dirac point caused by the same
 impurities that are responsible for the scattering of carriers at
 high density~\cite{kn:adam2007a}.  These two conceptually different
 approaches lead to strikingly different predictions for the
 conductivity at the Dirac point: The Boltzmann transport theory for
 Coulomb disorder predicts that increasing disorder {\em decreases}
 the conductivity, whereas the weak antilocalization picture has a
 conductivity that {\em increases} with increasing disorder strength.
 Given their vastly different starting points, it is not surprising
 that the two approaches disagree.

A direct comparison between the two approaches has not been possible,
mainly because the published predictions of the Boltzmann approach
include screening of the Coulomb disorder potential, whereas the fully
quantum-mechanical numerical calculations are for a non-interacting
model using Gaussian disorder.  Notwithstanding the fact that
screening and Coulomb scattering play crucial roles in transport of
real electrons through real 
graphene \cite{kn:footnote1}, the important question of the comparison between 
quantum and Boltzmann theories has remained
unanswered, even for the Gaussian disorder case.  It is the goal of
this work to provide such a comparison, thereby establishing the
bridge between these two widely used complementary theoretical
approaches to transport in graphene.

In what follows we consider non-interacting electrons at zero temperature
in a Gaussian correlated disorder potential that varies smoothly on
the scale of the graphene lattice spacing. This situation is described 
by the effective Hamiltonian
${\mathcal H} =  {\bf \sigma}\cdot {\bf p} + U({\bf r}),$
where $v$ is the Fermi velocity, ${\bf p}$ the (two-dimensional)
momentum, and $U({\bf r})$ a random Gaussian potential 
with correlation function
\begin{eqnarray}
  \langle U({\bf r}) U({\bf r'}) \rangle 
  &=& K_0 \frac{(\hbar v)^2}{2 \pi \xi^2} e^{-|{\bf r} - {\bf
  r'}|^2/2\xi^2},
\label{Eq:Hamilt}
\end{eqnarray}
where $\xi$ is the correlation length and
$K_0$ is a dimensionless parameter that parameterizes its
magnitude. Typical experimental conditions correspond to $K_0$ between
$1$ and $3$ \cite{kn:tan2007,kn:adam2008c}. 
We numerically solve the full 
quantum problem for a sample of finite size $L \gg \xi$, 
starting from 
intrinsic graphene where quantum coherence effects 
dominate, to high doping where quantum effects are 
a small correction to the conductivity \cite{kn:footnote2}.  We have compared
  the numerical results to predictions of the Boltzmann theory, its
  self-consistent modification of Ref.~\onlinecite{kn:adam2007a}, and
  weak antilocalization corrections, for a range of disorder strength
  $K_0$ and carrier densities $n$.

Our main conclusions, to be supported by the material below, are: (i)
Away from the Dirac point, both the Boltzmann theory and the full
quantum solution agree to leading order, $\sigma~ \propto n^{3/2}$,
with deviations only in terms of order $n^{1/2}$ and smaller. This
validates the assumptions of both theories, {\em i.e.}, the Born
approximation for the Boltzmann approach and use of a finite sample
size in the numerical approach.  (ii) At the Dirac point, the quantum
conductivity increases with increasing disorder strength $K_0$; for
$K_0 \gg 1$ the increase is compatible with the self-consistent
Boltzmann theory.  (iii) As a function of carrier density, the quantum
conductivity, but not the Boltzmann conductivity, shows a sharp
reduction at the Dirac point, which is most pronounced for $K_0 \sim
1$; the conductivity becomes proportional to $1/K_0$ away from the
Dirac point; (iv) The numerical quantum results are consistent with $d
\sigma/d\ln L = 4 e^2/\pi h$ for $\sigma \gtrsim 4 e^2/h$, {\it
irrespective of the carrier density $n$}, consistent with the weak
antilocalization theory. Our numerical calculations provide a
quantitative description of the full crossover between the quantum and
semiclassical transport regimes, for which no analytical theory is
available.

The Boltzmann conductivity corresponding to the model 
of Eq. \ref{Eq:Hamilt} is calculated using 
the relation $\sigma = e^2 \nu v_F^2 \tau/\hbar$, where $\nu = 4 k_F/
\pi \hbar v_F$ is the density of states and the elastic 
relaxation time $\tau$ is given by
\begin{eqnarray}
  \frac{1}{\tau} &=& 
  \int \frac{d\vq d\vr}{4 \pi \hbar} (1 - \cos^2 \theta_{\vq})
  \delta(k_F-q) 
  \langle U(0) U(\vr) \rangle e^{i \vq \cdot \vr}, \nonumber
  \label{eq:tau}
\end{eqnarray}
where $\theta_{\vq}$ parameterizes the direction of
$\vq$, so that
\begin{eqnarray}
  \sigma_{B} &=& \frac{4 e^2}{h} \frac{\pi n \xi^2
  e^{\pi n \xi^2}}{K_0 I_1(\pi n \xi^2)} \nonumber \\
  &=& \frac{2 \sqrt{\pi} e^2}{K_0 h} \left[ (2 \pi n \xi^2)^{3/2}
  + {\cal O}(n \xi^2)^{1/2} \right],
  \label{eq:sigmaB}
\end{eqnarray}
with the carrier density $n = k_F^2/\pi$.  The leading term
for large density can also be obtained considering the classical
diffusion of a particle undergoing small-angle deflections from the
random potential $U$ \cite{kn:footnote3}. 
The weak antilocalization correction to
the conductivity is \cite{kn:suzuura2002}
\begin{equation}
  \delta \sigma (L, \ell) 
  = \frac{4 e^2}{\pi h}  \ln \left( L/\ell \right),
\label{Eq:WAL}
\end{equation}      
where $\ell$ is the transport
mean free path. In the Boltzmann theory, $\ell$ can 
be obtained from the relation $\sigma_B = 2 (e^2/h) k_F \ell$.
A self-consistent modification of the Boltzmann theory was proposed in Ref.\
\onlinecite{kn:adam2007a} in order to describe transport near the
Dirac point $n=0$. For our Gaussian model of disorder, this modification 
involves replacing the carrier density by a ``self-consistent''
carrier density $n^* = \pi^{-1} (\varepsilon_F^*/\hbar
v_F)^2$, where $\varepsilon_F^{*2} = \langle (\varepsilon_F + U)^2
\rangle$ \cite{kn:footnote4}. We then find $n^* =
|n| + K_0/2 \pi^2 \xi^2$, and the self-consistent prediction for the
conductivity is given by Eq.\ (\ref{eq:sigmaB}) above with $n$
replaced by the self-consistent density $n^*$.

\begin{figure}
\bigskip
\includegraphics[scale=0.3]{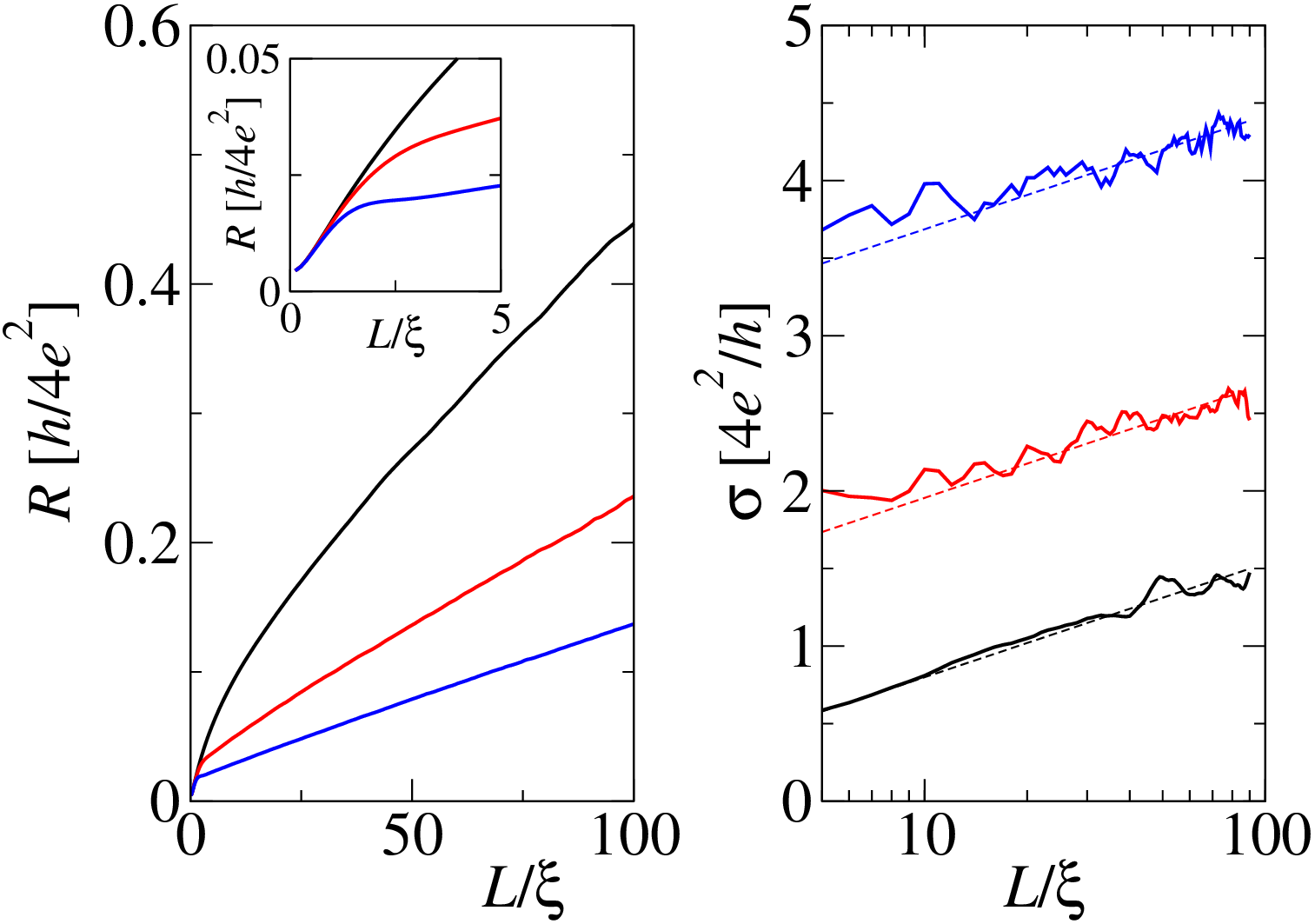}
\caption{\label{fig:fig1} (Color online) Resistance,
$R = 1/G$ (left) and conductivity $\sigma$ obtained using Eq.\
  (\ref{eq:condG}) (right), as a function of sample length $L$.
The three curves shown are for $W/\xi=200$,
$K_0=2$ and $\pi n \xi^2=0$, $0.25$, and
$1$ [from top to bottom (bottom to top) in left (right) panel].
Dashed lines in the right panel show d$\sigma/d \ln L = 4 e^2/\pi h$. 
The inset in the left panel shows the crossover 
to diffusive transport ($L \gg \xi$).}
\end{figure}

In the numerical calculation we consider a graphene strip of 
dimensions $L\times
W$ with $W, L \gg \xi$, connected to a highly doped graphene 
regions on both ends. Following the method described in
Ref.~\cite{kn:bardarson2007}, we calculate the conductance $G$ of the
graphene strip. The conductivity $\sigma$ is then obtained using the
relation
\begin{equation}
  \sigma = \left[ W \frac{dR}{dL} \right]^{-1}, \ \ R = 1/G.
  \label{eq:condG}
\end{equation}
We verify that our results do not depend on the
real-space discretization in the longitudinal direction, the cut-off 
of the transverse momentum (see Ref.\ \onlinecite{kn:bardarson2007}
for details), and the aspect ratio $W/L$. 
Extracting the conductivity using (\ref{eq:condG}) is different from
Ref.\ \onlinecite{kn:bardarson2007}, where the conductivity was 
identified with $LG/W$. The advantage of Eq.\ (\ref{eq:condG}) is that 
it eliminates the effect of an additive resistance from a region of
ballistic transport adjacent to the contacts to the source and drain
reservoirs and, hence, gives
accurate conductivities for smaller samples sizes than the
identification of $\sigma$ and $LG/W$. Our procedure is illustrated 
in Fig.\ \ref{fig:fig1}, where we show typical quantum numerical results for the resistance $R =
1/G$ and the conductivity $\sigma(L)$ defined through 
Eq.\ (\ref{eq:condG}).

\begin{figure}
\bigskip
\includegraphics[scale=0.35]{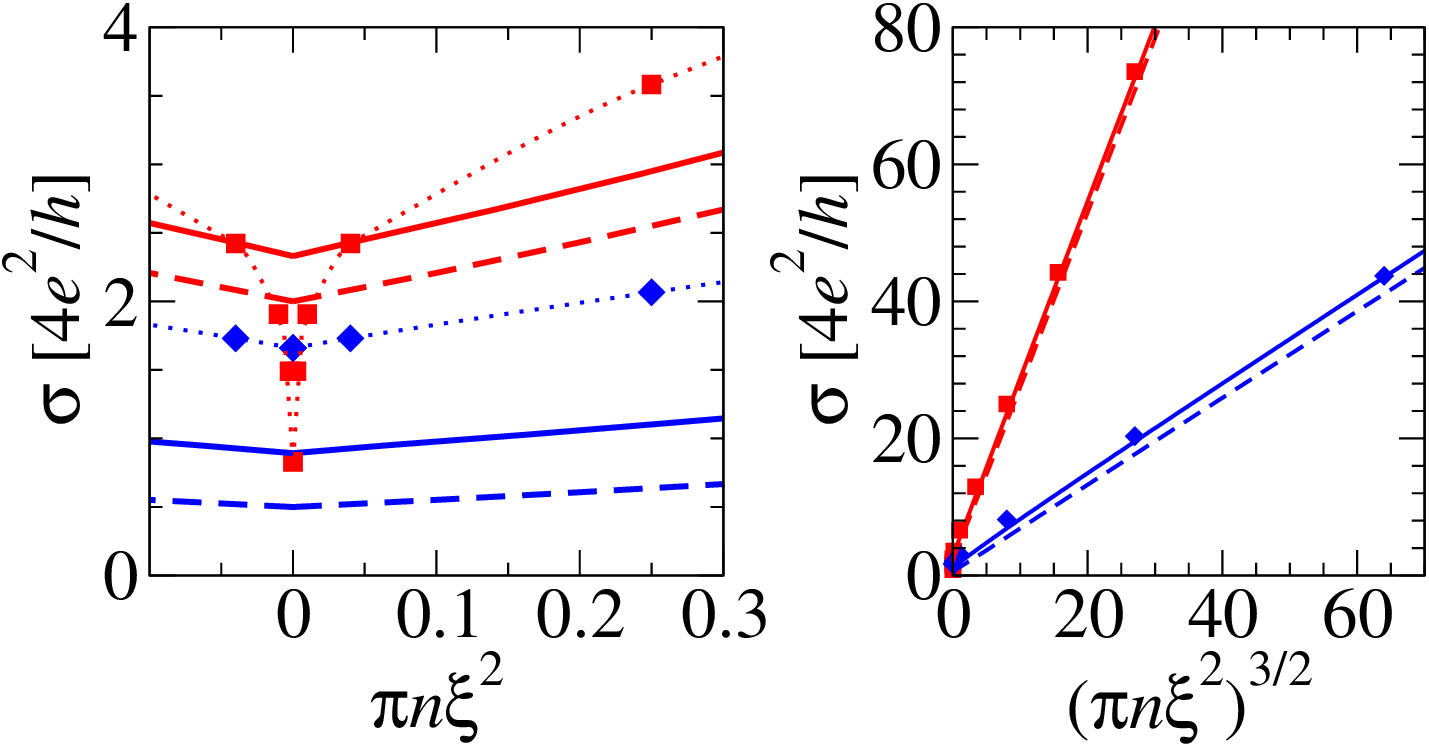}
\caption{\label{fig:fig2} (Color online) 
Conductivity $\sigma$ versus carrier density $n$. The left panel shows
the low-density behavior near the Dirac point, the right panel shows
the high-density behavior. Data points are from numerical simulation
with $K_0 = 1$ (squares) and $K_0=4$ (diamonds) with dotted lines 
in the left panel as a guide to the eye.  The dashed curves show the 
predictions of the Boltzmann theory, and the solid lines show the 
self-consistent Boltzmann result.}
\end{figure}

\begin{figure}
\bigskip
\includegraphics[scale=0.35]{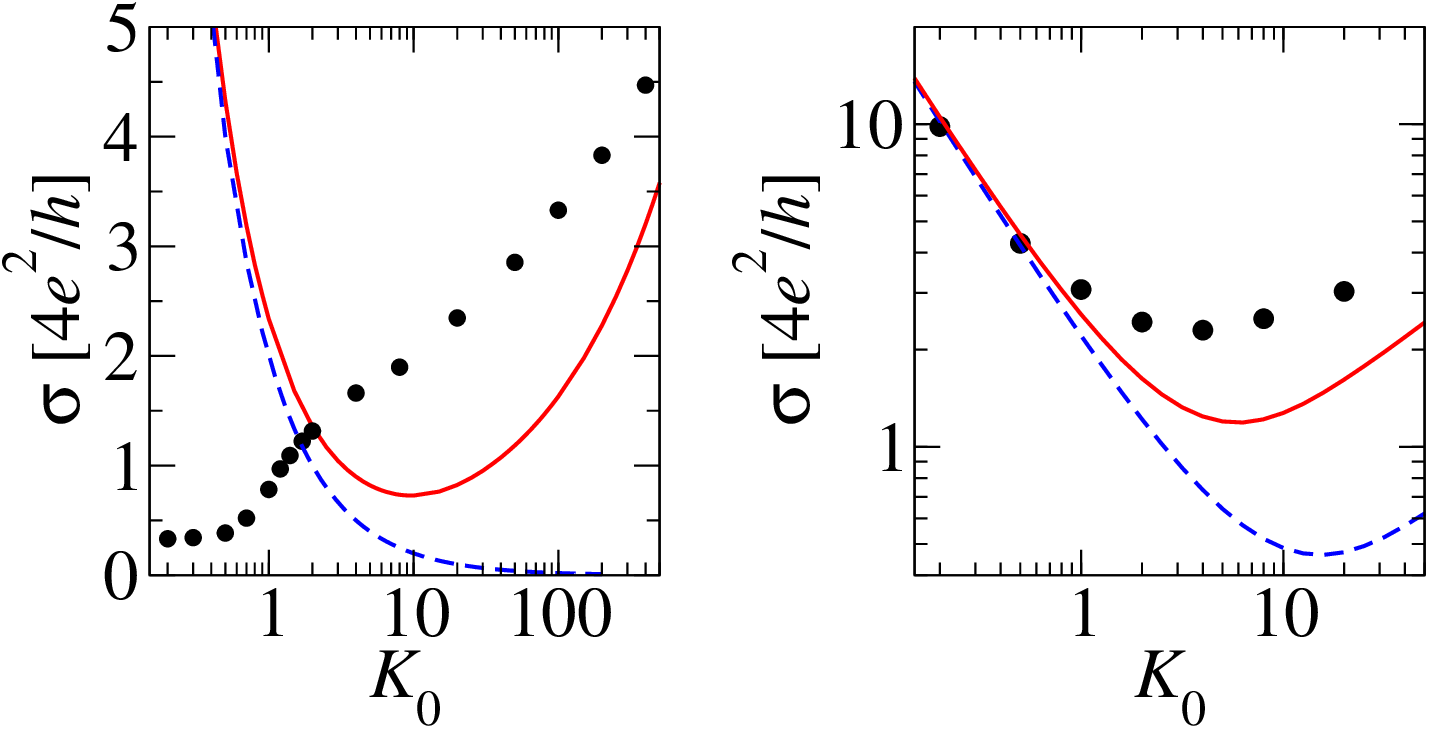}
\caption{\label{fig:fig3}
(Color online) Conductivity $\sigma$ versus disorder strength at the
  Dirac point (left) and at carrier density $ \pi n=K_0/(\pi \xi)^2$, 
  corresponding to the edge of
  the minimum conductivity plateau of Ref.\
  \onlinecite{kn:adam2007a} (right). 
  Data points are
  from the numerical calculation for $L = 50 \xi$ and the (solid)
  dashed curves represent the (self-consistent) Boltzmann theory.}
\end{figure}

We restrict the analysis of our numerical data to samples with length
$L \gtrsim \ell$. (In semiclassical transport, this is the regime
where electron motion is diffusive.) According to the Boltzmann
theory, $\ell \sim \xi^3 n/K_0$. In the diffusive regime, the quantum
conductivity $\sigma$ has a weak dependence on the sample length $L$
because of weak antilocalization. In order to compare numerical data
at different $K_0$ or $n$, we use two different procedures. 
In Figs.\ \ref{fig:fig2} and \ref{fig:fig3}, we 
compare conductivities at a reference
sample length $L=50 \xi$, which is well inside the diffusive regime if
$K_0 \gtrsim 1$.  For this sample size and beyond, the $L$-dependence
of the conductivity was consistent with the theoretical expectation $d
\sigma/d\ln L = 4 e^2/\pi h$ of weak antilocalization theory for all
carrier densities if $K_0 \gtrsim 1$.
(Representative data for $K_0=2$ shown in Fig.\
\ref{fig:fig1}.)  At the Dirac point, this size-dependence of the
conductivity was previously observed in
Refs.~\onlinecite{kn:bardarson2007,kn:nomura2007c,kn:tworzydlo2008}.

Figure \ref{fig:fig2} shows the conductivity $\sigma(L = 50 \xi)$ 
versus the carrier density $n$ for two values of $K_0$. For $n \xi^2 \gtrsim 1$
the conductivity is well described by the asymptotic behavior of Eq.\
(\ref{eq:sigmaB}), the dominant correction term being proportional to
$n^{1/2}$. Replacing the carrier density $n$ by the self-consistent
carrier density $n^*$ (solid lines) further improves the
agreement. For small densities, the quantum conductivity shows a sharp
minimum at the Dirac point $n=0$, which is most pronounced for small
disorder strengths. Such a dip is not present in either the Boltzmann
theory or its self-consistent modification. 
Since in the quantum
theory $\sigma$ increases with increasing $K_0$ at the Dirac point but
decreases with increasing $K_0$ away from the Dirac point, the quantum
$\sigma$ vs.\ $n$ curves of different $K_0$ cross somewhere in the
region $0 < \pi n \xi^2 \lesssim 1$ for the parameter range we
consider. This reversal in behavior was previously noted by Lewenkopf
{\em et al.} in numerical simulations of a tight-binding model
\cite{kn:lewenkopf2008}, although the numerical data of Ref.\
\onlinecite{kn:lewenkopf2008} do not allow a conclusion to be made
about large carrier density.  The agreement at high carrier density 
between the quantum and the Boltzmann theory is an important 
new result of this work.

In Fig.\ \ref{fig:fig3}, we address the conductivity as a function of
disorder, comparing the quantum and Boltzmann theories. Motivated by
prediction of Ref.\ \onlinecite{kn:adam2007a} that the $\sigma$ vs.\
$n$ curve exhibits a plateau of width $\sim K_0/\xi^2$ near the Dirac
point, we consider the conductivity at the Dirac point $n=0$ (left panel)
as well as near the edge of the proposed 
plateau, at $ \pi n = K_0/(\pi \xi)^2$ (right panel) \cite{kn:footnote5}. The numerical calculations at the plateau edge
are in good qualitative agreement with the
self-consistent Boltzmann theory.
At the Diract point, however, $\sigma$ is found to increase 
with $K_0$
for the entire parameter range we consider, which differs from the
prediction of the Boltzmann theory \cite{kn:footnote6} and the 
self-consistent Boltzmann
theory. The former predicts $\sigma = 8 e^2/ K_0 h$ at the Dirac
point, whereas the latter deviates from this prediction for $K_0 \sim
1$, reaches a minimum at $K_0 \approx 9.71$, and crosses over to the
asymptotic dependence $\sigma \sim 2 e^2K_0^{1/2}/\pi h$ for $K_0 \gg
10$. At large $K_0$ the numerical data follow the trend of the
self-consistent theory, although we cannot confirm the
asymptotic dependence $\propto K_0^{1/2}$ from the parameter
range studied in our simulations. Upon reducing $K_0$ below unity,
the conductivity first decreases sharply, consistent with a
renormalization of
the mean free path $\ell$ for $K_0 \lesssim
1$~\cite{kn:aleiner2006,kn:schuessler2008}. Upon reducing $K_0$
further, the Dirac point conductivity 
saturates at the ballistic value $\sigma = 4 e^2/\pi h$.

\begin{figure}
\bigskip
\includegraphics[scale=0.35]{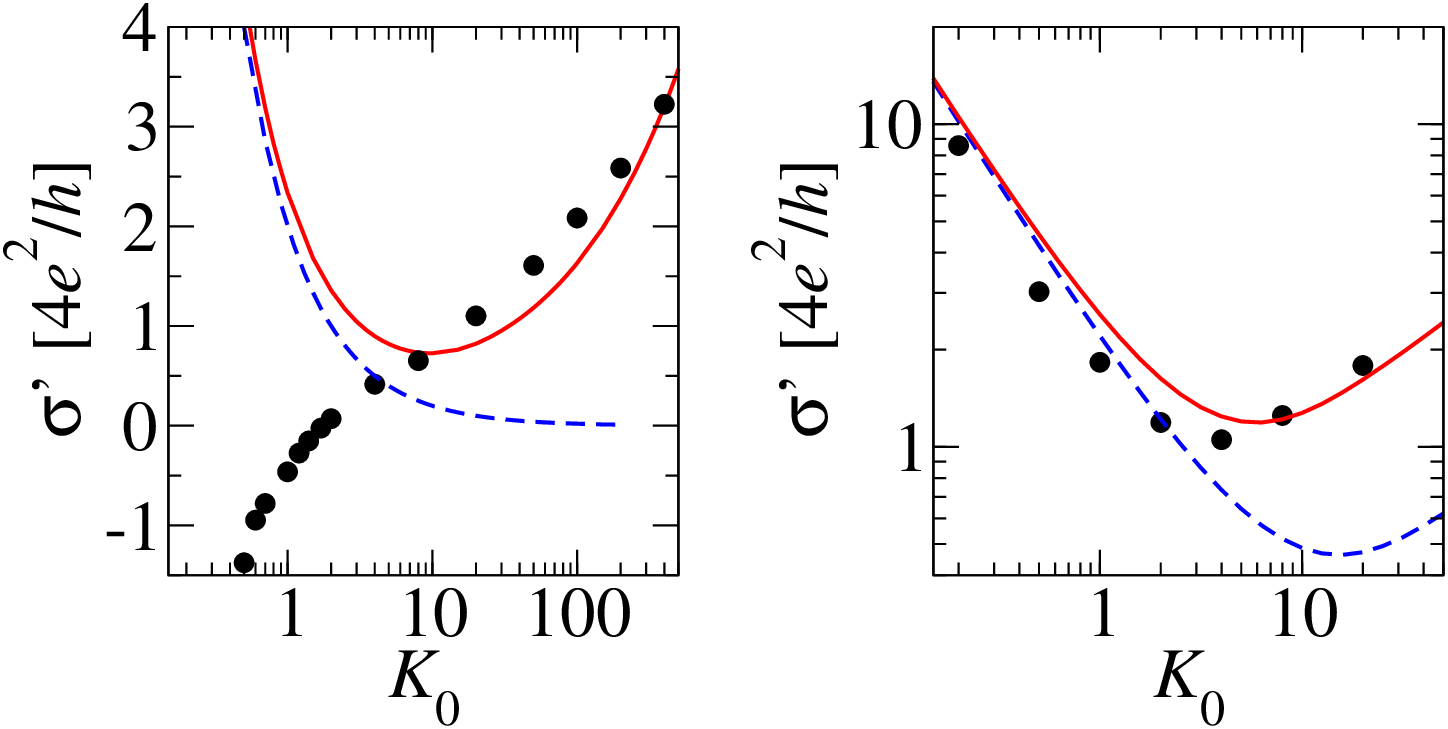}
\caption{\label{fig:fig4} (Color online) Same as Fig.\ \ref{fig:fig3},
  but for $\sigma' = \lim_{L \to \infty} [\sigma(L) -
\pi^{-1} \ln(L/\xi)]$.}
\end{figure}

 The system-size-dependent weak antilocalization correction
(\ref{Eq:WAL}) is included in Figs.\
\ref{fig:fig2} and \ref{fig:fig3}, which show $\sigma$ at
the reference length $L = 50 \xi$. In Fig.\ \ref{fig:fig4} we
subtract the $L$-dependent logarithmic increase and
show the $K_0$ dependence of $\sigma' = \lim_{L \to \infty} 
[\sigma(L) -
\pi^{-1} \ln(L/\xi)]$ \cite{kn:footnote7}. 
Subtracting weak antilocalization significantly improves
the agreement with the self-consistent theory at large $K_0$. Unlike
the conductivity at the reference length $L=50\xi$, which
saturates at $4 e^2/\pi \hbar$ for small $K_0$, 
$\sigma'$ continues to decrease without bounds if $K_0$ is lowered.
 
The increase of $\sigma$ with $K_0$ at the Dirac point for weak
disorder is markedly different from the prediction of the Boltzmann
theory. A key assumption of this theory and its
self-consistent modification is that the graphene electron liquid can
be mapped to an essentially homogeneous system with an effective
carrier density $n^*$ equal to the rms of a fluctuating ``local'' 
density determined by the random potential $U$. This assumption 
becomes questionable at
the Dirac point, where the electron liquid is broken up in puddles of
electron-like and hole-like regions.
 At weak disorder, $K_0 \ll 1$, quantum fluctuations spread the
carriers over many puddles and the concept of a local carrier density 
becomes problematic. It is in this regime that the difference between the
quantum and Boltzmann calculations are, as expected, most pronounced.

The Gaussian random potential used here is the potential-of-choice for
comparisons of analytical theories and numerical simulations.  Yet, it
differs in essential ways from the random potential in realistic
graphene samples that do not have Gaussian statistics, since it is
likely caused by charged impurities in the substrate with a typical distance
$d$ from the graphene sheet smaller than the spacing between
impurities.  Still, it may be possible to extract equivalent
parameters $K_0$ and $\xi$ from a realistic random potential (see
Ref.\ \onlinecite{kn:adam2008c}), implying that the sharp dip in
conductivity predicted in the quantum theory would occur in a window
$n \sim 5 \times~10^{10}~{\rm cm}^{-2}$ around the Dirac 
point~\cite{kn:footnote8}.  This feature has not been observed in
experiments
\cite{kn:novoselov2004,kn:tan2007,kn:chen2008,kn:jang2008}.  Reasons
why the dip has not been seen could be a suppression of quantum
coherence by finite temperature effects or
rippling~\cite{kn:morozov2006} of the graphene sheet, or long-range
fluctuations of the mean carrier density which effectively smear the
feature near $n=0$ \cite{kn:footnote9}.

We thank C.\ Beenakker for comments on the manuscript.
SA thanks the Aspen Center for
Physics for its hospitality where some of this work was completed.  
This work is supported by US-ONR,
NSF-NRI-SWAN, the Packard Foundation, and by the NSF under grant no.\
DMR 0705476.

\vspace{-0.2in}

\begin{thebibliography}{37}
\bibitem[{kn:({\natexlab{b}})}]{kn:footnote2}
Quantum effects are a small correction to the
conductivity only if the carrier density $n$ is increased at fixed
sample size $L$. This is the experimentally relevant limit. If the
limit $L \to \infty $ is taken at fixed $n$, quantum effects dominate,
see Eq.\ (\ref {Eq:WAL}). We are therefore not considering the
conceptually simple question of how quantum transport becomes
classical as the phase coherence length decreases, but the more
interesting question of how this quantum-Boltzmann crossover depends
on the carrier density and disorder strength.
\bibitem[{kn:({\natexlab{c}})}]{kn:footnote3}
The Boltzmann conductivity differs from that of
the self-consistent Born approximation (SCBA) if $K_0$ is not
small. The origin of the discrepancy is the predominance of forward
scattering if $n \xi ^2 \gg 1$ and the breakdown of the non-crossing
approximation in this case.
\bibitem[{kn:({\natexlab{e}})}]{kn:footnote5}
We have also considered a model
where we replace the random potential $U$ with its absolute value
$|U|$, which gives results similar to those of the right panel of
Fig.\ \ref {fig:fig3}.
\end{thebibliography}

\end{document}